\documentclass[journal,twoside,web]{ieeecolor}
\usepackage{generic}
\usepackage{amsmath,amssymb,amsfonts}
\usepackage{algorithmic}
\usepackage{graphicx}
\usepackage{tabularx}
\usepackage{subcaption}
\usepackage{circledsteps}
\def\BibTeX{{\rm B\kern-.05em{\sc i\kern-.025em b}\kern-.08em
    T\kern-.1667em\lower.7ex\hbox{E}\kern-.125emX}}
\usepackage{textcomp}
\usepackage[dvipsnames]{xcolor}
\usepackage{siunitx}
\usepackage{multirow}
\usepackage[hidelinks]{hyperref}
\usepackage{flushend}
\usepackage[noabbrev,capitalise]{cleveref}
\Crefname{figure}{Fig.}{Fig.}
\usepackage[
        defernumbers=true,
        sortcites,
        backend=biber,
        bibencoding=utf8,
        natbib=true,
        hyperref=true,
        backref=false,
        urldate=long,
        style=ieee,
        isbn=false,
        dashed=false,
        url=true,
        eprint=false,
        maxnames=1,
        sorting=none
    ]{biblatex}
\usepackage{textcomp}
\def\BibTeX{{\rm B\kern-.05em{\sc i\kern-.025em b}\kern-.08em
    T\kern-.1667em\lower.7ex\hbox{E}\kern-.125emX}}
\begin{document}
\title{System--Technology Co-Evaluation of \\ 
A7 CFET and A10 NSFET Technologies \\ 
from Cell Parasitics to Chip Reliability}
\author{Mahdi Benkhelifa, Leon Mayr, Hadi Nour Eddine, Andrea Padovani, \IEEEmembership{Member, IEEE}, \\ Luca Larcher, and Hussam Amrouch, \IEEEmembership{Member, IEEE}
\thanks{Received XX XXXX 2026. (Corresponding author: Hussam Amrouch.)}
\thanks{M. Benkhelifa, L. Mayr, H. N. Eddine, and H. Amrouch are with the Technical University of Munich, TUM School of Computation, Information and Technology, Chair of AI Processor Design, Munich Institute of Robotics and Machine Intelligence, Munich, Germany (e-mail: m.benkhelifa@tum.de; leon.h.mayr@tum.de; hadi.nour@tum.de; amrouch@tum.de).}
\thanks{A. Padovani is with the Department of Science and Methods for Engineering, University of Modena and Reggio \\ Emilia, Reggio Emilia, Italy (e-mail: andrea.padovani@unimore.it).}
\thanks{L. Larcher is with Applied Materials, CA, USA (e-mail: Luca\_Larcher@amat.com).}%
}
\maketitle

\begin{abstract}
Complementary FETs (CFETs) extend nanosheet FET (NSFET) scaling by vertically stacking n- and p-type gate-all-around (GAA) devices, thereby shrinking standard-cell area. The performance gain, however, cannot be assessed from device metrics alone, as CFET layouts also introduce larger cell-level parasitic resistance and capacitance (RC). In this work, we present a physics-based thermal- and aging-aware system--technology co-evaluation (STCO) flow to assess parasitic RCs in A7 CFET and A10 NSFET technology nodes. Our flow links calibrated device models, optimized standard-cell generation, automated GDS-to-TCAD conversion enabling accurate 3D parasitic RC extraction, full RTL-to-GDS implementation for an AI accelerator, multiphysics thermal analysis, and physics-based bias temperature instability (BTI) aging evaluation. Using the same device model for both technologies, we can isolate the impact of parasitic RCs and design at different levels of the design flow. The results of the AI accelerator design demonstrate that the A7 CFET reduces the chip area by \qty{24.7}{\percent} and the total wire length by \qty{12}{\percent}, improving the area efficiency \(\mathbf{\mathrm{TOPS}/\mathrm{mm^2}}\) by \qty{74}{\percent} relative to the baseline of the A10 NSFET. Under iso-frequency operation, results reveal that CFET voltage scaling reduces power by \qty{68}{\percent} and lowers power density from \qty{148}{\watt\per\centi\meter\squared} to \qty{55}{\watt\per\centi\meter\squared}, which reduces the chip's temperature from \qty{125}{\celsius} down to merely \qty{62}{\celsius}. The resulting reduction in stress temperature suppresses 10-year BTI-induced degradation by \qty{39}{\percent}, reducing the required aging timing guardband by \qty{53}{\percent}.
\end{abstract}


\begin{figure}[t]
    \centering
    \includegraphics[width=0.5\textwidth]{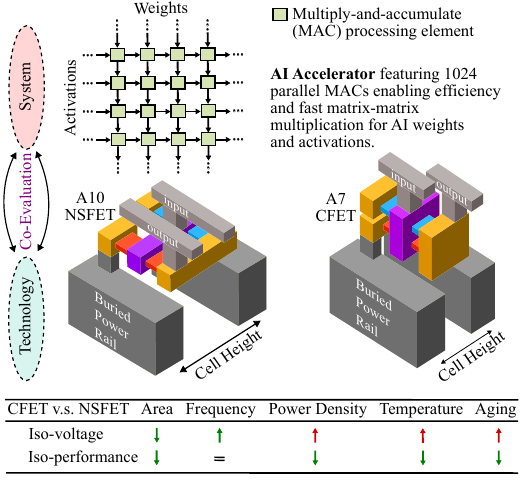}
    \caption{CFET vertically stacks n- and p-type nanosheets, reducing cell height and hence chip area. However, tighter 3D integration increases cell parasitics and hence switching power leading to higher power density, increased temperature, and accelerated aging. Nevertheless, under iso-performance operation, CFET achieves the target frequency at a lower voltage, reducing power, power density, temperature, and aging.}
    \label{fig:introduction_nsfet_cfet}
\end{figure}

\begin{figure*}[t]
\centerline{\includegraphics[width=\textwidth]{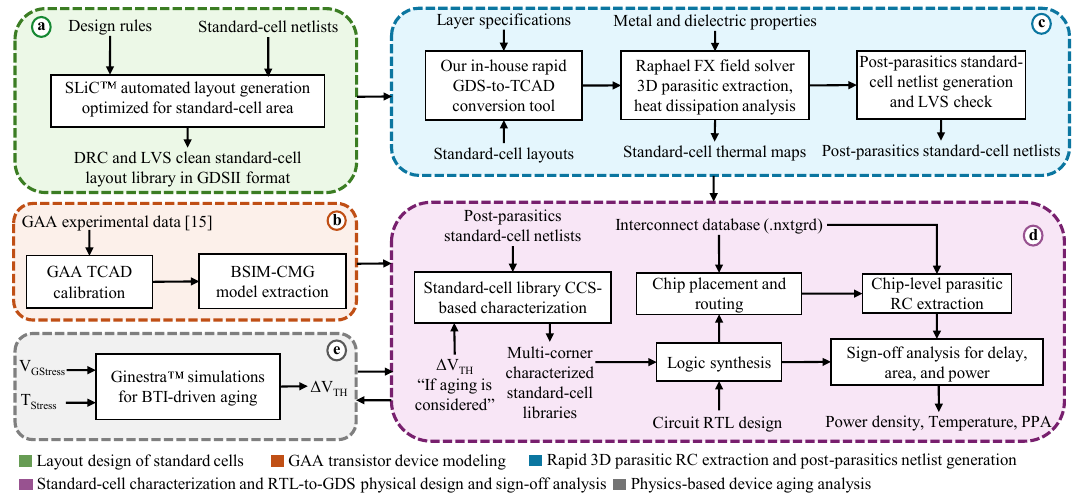}}
\caption{System–technology co-evaluation workflow. (a) SLiC\texttrademark{} enables automated, optimized standard-cell layout generation. (b) Transistor characteristics are calibrated to measurements, and a SPICE compact model is calibrated. (c) Our in-house GDS-to-TCAD framework generates 3D TCAD structures, from which Raphael FX extracts standard-cell parasitic RCs and 3D thermal maps. (d) Our digital design flow includes library characterization, RTL-to-GDS implementation, chip-level parasitic RC extraction, and industrial sign-off analysis to obtain PPA. Multiphysics simulations convert power density into chip-level temperature. (e) Physics-based BTI defect generation using Ginestra\texttrademark{} for accurate BTI-induced $\Delta V_{TH}$ analysis.}
\label{fig:STCO_flow}
\end{figure*}

\section{Introduction}
\label{sec:introduction}
\IEEEPARstart{N}{SFETs} are a key architecture for nodes beyond \qty{5}{\nano\meter}~\cite{nsfet_scaling_importance2}. Their vertically stacked channels, fully surrounded by the gate, provide superior electrostatic control while increasing effective channel width—and hence drive-current capability—within a given device
footprint~\cite{nsfet_control1,nsfet_drive}. However, further logic scaling is increasingly constrained by standard-cell layout rather than by transistor size. Lateral n-/p-device separation, contacted poly pitch (CPP), gate length, and local routing resources limit further reductions in cell area and hence logic density~\cite{scaling_limits,np_minspacing,cpp_scaling,scaling_lateral}. CFETs address these limitations by vertically stacking n- and p-type GAA transistors, enabling aggressive footprint reduction~\cite{cfet_scalability}.

The CFET density advantage is accompanied by increased parasitic and reliability challenges. Compact vertical integration introduces additional high-aspect-ratio interconnect resistance and substantial MOL parasitic capacitance, while the close proximity of the stacked devices enhances electrical and thermal coupling \cite{cfet_parasitics,cfet_parasitics2, cfet_reliability,cfet_reliability2}.
\begin{figure}[t]
    \centering
    \includegraphics[width=0.45\textwidth]{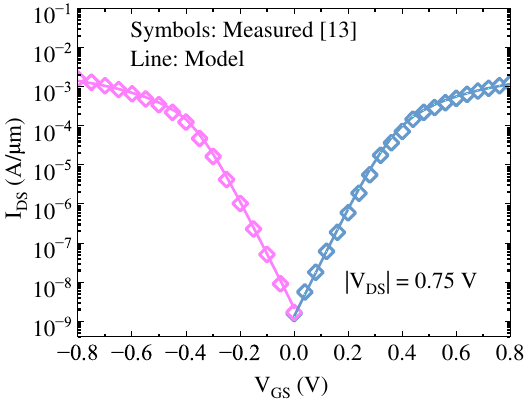}
    \caption{TCAD-calibrated model for n-type and p-type nanosheets. $I_{DS}–V_{GS}$ transfer characteristics show an excellent match with data measurements \cite{Liao_2023_TSMC_CFET}, which we calibrated against.}
    \label{fig:TCAD_calibartion}
\end{figure}
Aggressive interconnect dimensional scaling increases resistivity through enhanced surface and grain-boundary electron scattering, while BPR-based backside power delivery and dense local routing introduce additional MOL parasitics and increase the complexity of FEOL–MOL–BEOL parasitic interactions~\cite{route_resources_and_increased_resistivity,interconnect_scaling,bpr_parasitics}. These effects degrade standard-cell delay and transition time as well as increase dynamic power. At the same time, reduced chip area results in higher power density \cite{a5_vs_a10,power_density} and, hence, higher temperature, which in turn accelerates BTI-induced defect generation.

Therefore, CFET scaling must be carefully analyzed through a system--technology co-evaluation flow that propagates technology assumptions into layout, parasitics, timing, thermal behavior, and even aging. Isolated device or cell metrics are insufficient because cell-level RC penalties and chip-level routing benefits can act in opposite directions. Reliability and performance consequences of the parasitic RCs that result from the vertical stacking in the A7 node compared to the A10 node are summarized in \cref{fig:introduction_nsfet_cfet}.

In our previous work~\cite{Benkhelifa_IRPS_2026}, we investigated the impact of cell-level parasitics in vertically stacked NSFET devices on critical-path delay using synthesized circuits and presented preliminary results on their implications for device self-heating. In this work, we evaluate how parasitics associated with vertical device stacking affect the PPA of placed-and-routed AI accelerator designs in the A7 CFET and A10 NSFET nodes, and ultimately their impact on device reliability. We show that, although the A7 CFET cell design incurs a measurable cell-level parasitic penalty, its greater layout density reduces total interconnect length at the routed-chip level, enabling higher operating frequencies. At iso-frequency, this performance advantage allows the CFET design to operate at a substantially lower supply voltage, reducing self-heating and, consequently, aging-induced performance degradation.


\section{system-technology co-evaluation flow}

Fig.~\ref{fig:STCO_flow} illustrates our system-technology co-evaluation flow, which includes a calibrated transistor model, automated standard-cell layout generation, library characterization, a full RTL-to-GDS flow, and multiphysics aging simulation.

\subsection{Our Device-to-System Flow}

First, SLiC\texttrademark{}~\cite{slic} automatically generates area-optimized GDSII layouts for a 75-cell standard-cell library using standard-cell netlists~\cite{gt3,asap7} and technology design rules.

\begin{table}[t]
\centering
\caption{Employed design parameters and assumptions of A7 and A10 technologies \cite{IRDS2024MoreMoore,n2_imec}.}
\label{tab:technology_design_parameters}
\setlength{\tabcolsep}{4pt} 
\begin{tabular}{|l|c|c|c|c|c|c|c|}
\hline
\multirow{2}{*}{\textbf{Layers}} & \multicolumn{2}{c|}{\textbf{width} [\unit{\nano\meter}]} & \multicolumn{2}{c|}{\textbf{spacing} [\unit{\nano\meter}]} & \textbf{line resist.} & \multicolumn{2}{c|}{\textbf{permittivity}} \\ \cline{2-5} \cline{7-8}
& \textbf{A10} & \textbf{A7} & \textbf{A10} & \textbf{A7} & [\unit{\ohm\per\micro\meter}] & \textbf{A10} & \textbf{A7}  \\ \hline
M0, M2, M3&10&10&10&8&889&2.95&2.375\\ 
M1&14&14&32&30&375&2.95&2.375\\ 
M4, M5&23&23&13&13&134&2.475&2.475\\
M6 - M9  & 40 & 40 & 40 & 40 & 16 & 2.475 &2.475\\ 
M10 & 80 & 80 & 80 & 80 & 3 & 2.475 &2.475\\ 
\hline
\end{tabular}
\end{table}

\begin{figure}[t]
    \centering
    \includegraphics[
  trim=104.05mm 65.63mm 104.05mm 65.63mm,
  clip,
  width=0.5\textwidth
]{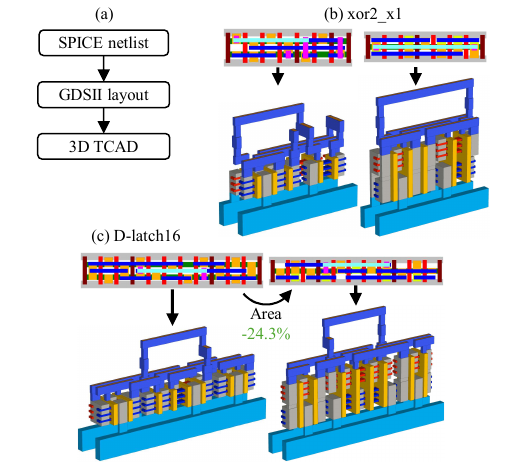}
    \caption{GDS-to-TCAD fully automated conversion framework. (a) SLiC™ generates optimized GDSII standard-cell layouts from SPICE netlists, which are automatically converted into 3D TCAD structures using our in-house framework. Example of layouts and corresponding 3D TCAD structures of X1-drive (a) XOR and (b) D-latch cells. The A7 cells achieve an average area reduction of \qty{24.3}{\percent}.}
    \label{fig:GDS_to_TCAD_framework}
\end{figure}

At the device level, a TCAD GAA nanosheet is calibrated~\cite{Shahin_2025_GAA_calibration} against measured $I_D$--$V_G$ data~\cite{Liao_2023_TSMC_CFET}. The calibrated device replicates the reported measurements with high accuracy, as shown in Fig.~\ref{fig:TCAD_calibartion}. BSIM-CMG compact models for n- and p-type devices are calibrated and used in cell-level SPICE simulations. We intentionally used the same calibrated device in both CFET and NSFET to isolate the impact of parasitics that result from vertical stacking and design-rule scaling.

Our in-house GDS-to-TCAD conversion framework automatically constructs full 3D TCAD standard-cell structures from the generated optimized layouts. The framework uses layer geometry and material specifications to enable accurate 3D field-solver-based parasitic RC extraction using Synopsys Raphael FX~\cite{Raphael}. Finally, post-parasitics standard-cell netlists are generated, and LVS checks are performed.

\begin{figure}[t]
    \centering
    \includegraphics[width=0.45\textwidth]{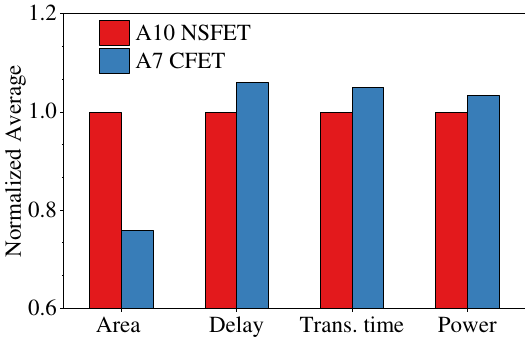}
    \caption{Normalized area, delay, transition time, and power of A10 NSFET- and A7 CFET-based standard cells. Values are averaged across all standard cells and normalized to the NSFET as a baseline. Higher parasitic RCs in CFET increase cell-level delay, transition time, and power.}
    \label{fig:results_standard_cells}
\end{figure}

The generated post-parasitics netlists are then characterized using the calibrated devices under the industry-standard composite-current-source (CCS) methodology across multiple voltage-temperature corners. The switching behavior of standard cells is modeled as a time-dependent output current waveform, resulting in an accurate description of delay, output transition, and power consumption at the cell level. Additionally, the liberty file generated in this step contains information on the input pin capacitance for each cell and pin, which can be used for technology comparison.

The characterized libraries are then used to synthesize the RTL design of a 64-bit RISC-V processor \cite{asanovic_2016_rocket}, and an in-house 32$\times$32 multiply-and-accumulate (MAC) systolic array AI accelerator. The synthesized AI accelerator designs are deployed in physical design, resulting in DRC-clean chip layouts. Chip-level parasitic extraction is performed to enable post-physical-design sign-off analysis and obtain the chip PPA.
Finally, the chip-level PPA results are then used to derive the temperature and voltage stress conditions for subsequent device-level reliability analysis.

\begin{figure}[t]
    \centering
    \includegraphics[width=0.45\textwidth]{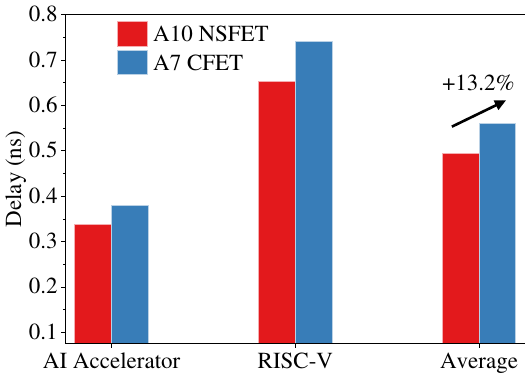}
    \caption{Delay comparison before chip physical design (\textbf{i.e., at the synthesis level}). RTL designs of the AI accelerator and a RISC-V processor are synthesized with the A10 and A7 cell libraries. CFET shows \qty{13.2}{\percent} larger delay due to higher cell parasitics.}
    \label{fig:results_synth}
    \vspace{-3mm}
\end{figure}

\subsection{BTI-driven Aging Evaluation} 

To compare the impact of parasitics on BTI-driven aging in NSFET and CFET devices, the chip power density from the sign-off analysis is used to run a package-level multiphysics ANSYS simulation to determine the maximum chip temperature under the given operating voltage. Ginestra™ \cite{Ginestra} simulations use the extracted temperature and the operating voltage as stress inputs to determine the 10-year threshold-voltage shift ($\Delta V_{TH}$) due to BTI aging. To evaluate the performance under aging, we re-characterize the standard-cell library using aged device characteristics and propagate the updated library through the RTL-to-GDS flow. Finally, sign-off analysis is performed to quantify the resulting impact of aging on system-level PPA.


\subsection{Technology Assumptions}

Technology assumptions for A10 NSFET and A7 CFET are derived from the IRDS 2024 roadmap and literature \cite{IRDS2024MoreMoore, n2_imec}. The resulting parameters are summarized in Table~\ref{tab:technology_design_parameters}. We apply node-to-node scaling assumptions for the FEOL up to M3, while keeping the BEOL from M4 to M10 identical for both technologies.
NSFET uses a \qty{46}{\nano\meter} CPP, a 4-track cell height, and a \qty{20}{\nano\meter} M0-pitch, whereas the A7 CFET uses a \qty{44}{\nano\meter} CPP, 3.5-track cell height, and a \qty{18}{\nano\meter} M0-pitch, reflecting tighter scaling in the CFET case. Line width and line resistance are identical for both NSFET and CFET for the entire BEOL stack. The spacer dielectric permittivity is scaled from \qty{3}{} to \qty{2.7}{}, the permittivty for M0-M4 is scaled from \qty{2.95}{} to \qty{2.375}{}, while M5-M10 are identically at \qty{2.475}{}, in alignment with the IRDS~\cite{IRDS2024MoreMoore}.
Fig.~\ref{fig:GDS_to_TCAD_framework} shows examples of the 3D TCAD structures of XOR and D-latch standard cells generated using our GDS-to-TCAD framework.

\begin{figure}[t]
    \centering
    \includegraphics[width=0.45\textwidth]{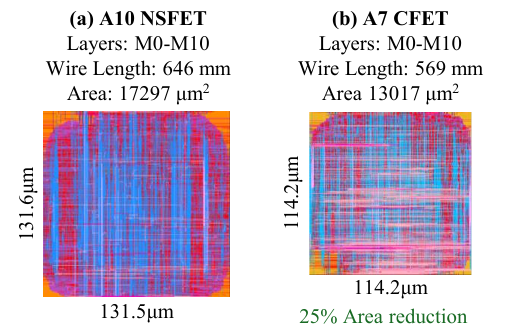}
    \caption{Full physical design showing GDS of AI accelerator using (a) A10 NSFET and (b) A7 CFET nodes. The CFET implementation reduces die area by \qty{25}{\percent} and shrinks total wire length by ~\qty{12}{\percent}, while still maintaining an equal area utilization of \qty{61.7}{\percent}.}
    \label{fig:results_PnR_layouts}
\end{figure}


\section{Results}

We evaluate the impact of parasitic RC components in NSFET and CFET technologies across multiple stages of the STCO flow. First, their effect on individual standard-cell performance is assessed. Circuit-level PPA is then evaluated after synthesis and again after physical design, enabling the impact of chip-level interconnect parasitics to be distinguished from cell-level parasitic effects. Finally, BTI-induced aging is analyzed at iso-frequency under the corresponding operating voltage and the obtained temperature conditions.

\subsection{Standard Cells}

 Fig.~\ref{fig:results_standard_cells} compares normalized area, delay, transition time, and power averaged across all $75$ sequential and combinational standard cells which we have created. A7 CFET reduces standard-cell area by \qty{24.3}{\percent} through p-/n-type transistor stacking. However, the compact 3D cell geometry increases the input pin capacitance by \qty{9.7}{\percent} on average across cells and input pins, despite the reduced dielectric permittivity. Additionally, super-vias increase parasitic resistance. Consequently, this leads to a \qty{6.1}{\percent} higher delay, a \qty{5.0}{\percent} longer transition time, and a \qty{3.4}{\percent} larger cell internal power. 
 
\begin{table}[t]
\centering
\caption{Iso-voltage (0.8V) post-physical-design evaluation of the AI accelerator designs. The A7 CFET design is faster than its A10 NSFET counterpart due to a \qty{12}{\percent} reduction in total wire length, which improves chip-level parasitics and results in higher TOPS. }
\label{table}
\setlength{\tabcolsep}{4pt}
\begin{tabular}{|l|c|c|l|}
\hline\centering
Metric& 
A10& 
A7&
Change\\
\hline
Frequency (\qty{}{\giga\hertz})& 
0.68& 
0.88&
\textcolor{green!50!black}{$+$\qty{30}{\percent}}\\
Power (\qty{}{\milli\watt})& 
12& 
21&
\textcolor{red}{$+$\qty{74.5}{\percent}}\\
Area (\qty{}{\micro\meter\squared})& 
17296& 
13017&
\textcolor{green!50!black}{$-$\qty{24.7}{\percent}}\\
Performance (TOPS)& 
1.39& 
1.81&
\textcolor{red}{$-$\qty{30}{\percent}}\\
Efficiency (TOPS/W)& 
115& 
86&
\textcolor{red}{$-$\qty{20.4}{\percent}}\\
Power Density (\qty{}{\watt\per\centi\meter\squared})& 
70& 
162&
\textcolor{red}{$+$\qty{132}{\percent}}\\
TOPS/mm$^2$& 
80& 
140&
\textcolor{green!50!black}{$+$\qty{28.1}{\percent}}\\
Figure of Merit (TOPS$^2$/mm$^2$*W)& 
9252& 
11958&
\textcolor{green!50!black}{$+$\qty{29.3}{\percent}}\\
\hline
\end{tabular}
\label{tab:results_MAC_table}
\end{table}


\begin{figure*}
\centering
\begin{subfigure}[]{0.45\textwidth} 
        \caption{}
        \centering
        \includegraphics[width=7.75cm]{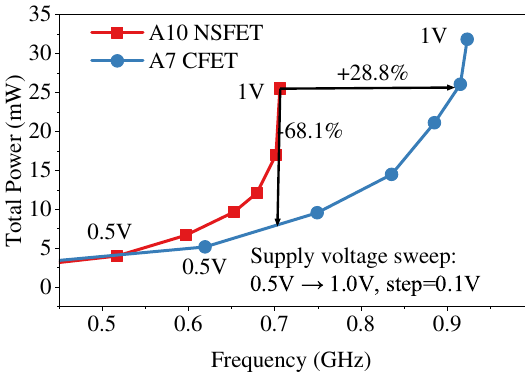}
    \label{fig:delay_power}
\end{subfigure}
\begin{subfigure}[]{0.45\textwidth} 
        \caption{}
        \centering
        \includegraphics[width=7.75cm]{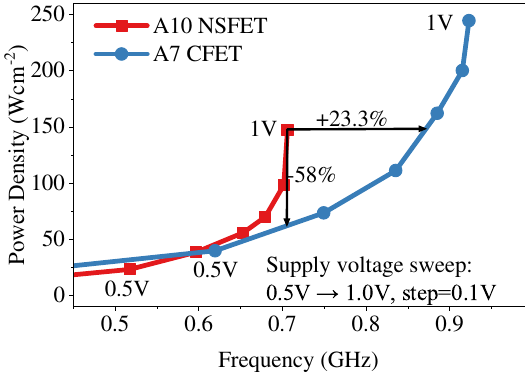}
    \label{fig:delay_powerdensity}
\end{subfigure}

    \caption{Power/performance tradeoffs of AI accelerator in A10 NSFET and A7 CFET. (a) The CFET design achieves iso-frequency at \qty{0.56}{\volt} vs. \qty{1}{\volt} for NSFET, reducing power by \qty{68}{\percent}. (b) Corresponding power-density scaling: NSFET reaches \qty{148}{\watt\per\centi\meter\squared} at \qty{1}{\volt}, while CFET reduces it to \qty{55}{\watt\per\centi\meter\squared} at \qty{0.56}{\volt}, leading to much lower temperature (Fig.~\ref{fig:full_chip_thermal}).}
\label{fig:scaling}
\vspace{-3mm}
\end{figure*}

\begin{figure}[t]
    \centering
    \includegraphics[width=0.45\textwidth]{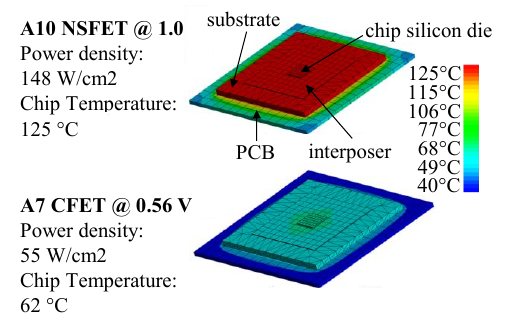}
    \caption{Power density and temperature of A10 NSFET vs. A7 CFET designs at iso-frequency (\qty{0.7}{\giga\hertz}). Chip temperature is obtained by mapping power densities (Fig.~\ref{fig:delay_powerdensity}) through chip-level multiphysics ANSYS simulations.}
    \label{fig:full_chip_thermal}
\end{figure}

\subsection{A10- and A7-based AI Accelerator Evaluation}

 \textbf{Pre-physical-design} timings obtained from the synthesized designs show that the A10 NSFET AI accelerator and RISC-V designs achieve delays of \qty{0.337}{\nano\second} and \qty{0.652}{\nano\second}, respectively. The A7 CFET implementations, on average, reduce the synthesized design area by \qty{24.7}{\percent}, while being \qty{13.2}{\percent} slower due to increased cell-level parasitics, which aligns well with the standard cell results (Fig.~\ref{fig:results_synth}). 
 

\begin{figure*}
\centering
\begin{subfigure}[]{0.45\textwidth} 
        \caption{}
        \centering
        \includegraphics[width=7.75cm]{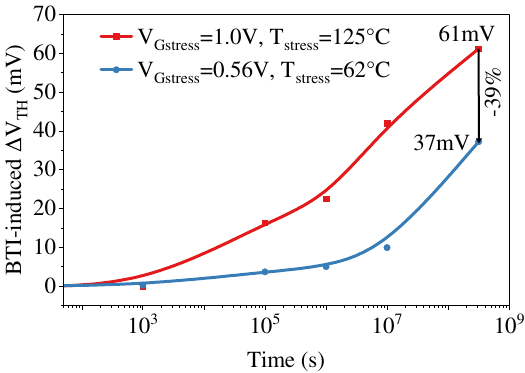}
    \label{fig:deltavth}
\end{subfigure}
\begin{subfigure}[]{0.45\textwidth} 
        \caption{}
        \centering
        \includegraphics[width=7.75cm]{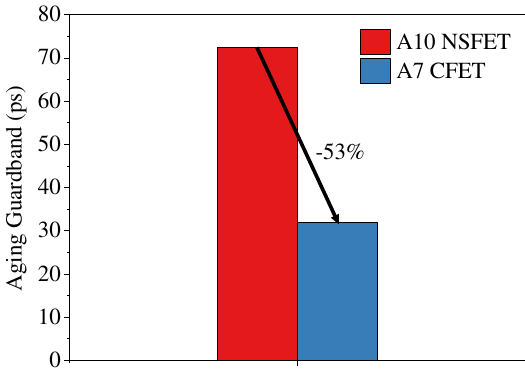}
    \label{fig:guardband}
\end{subfigure}

        \caption{Physics-based BTI aging analysis using Ginestra™. (a) After 10 years, $V_{TH}$ increases by \qty{61}{\milli\volt} at (\qty{1}{\volt}, \qty{125}{\celsius}) vs. \qty{37}{\milli\volt} at (\qty{0.56}{\volt}, \qty{62}{\celsius}). (b) Post-aging delay signoff analysis shows a \qty{53}{\percent} reduction in required aging timing guardband for the AI accelerator under CFET stress conditions. Results were obtained using aging-aware NSFET and CFET standard-cell libraries, which were characterized under the effects of aging-induced increases in $V_{TH}$.}
\label{fig:aging}
\vspace{-3mm}
\end{figure*}

\textbf{The physical implementations} of the AI accelerator chips in A10 NSFET and A7 CFET are shown in Fig.~\ref{fig:results_PnR_layouts}. The accelerator features a $32 \times 32$ MAC systolic array. Using the same M0--M10 routing stack and an identical area utilization of \qty{61.7}{\percent}, the CFET-based accelerator achieves a \qty{25}{\percent} smaller footprint than its NSFET-based counterpart owing to its reduced standard-cell height. The CFET implementation also reduces the total wire length by \qty{12}{\percent}, indicating more efficient routing and lower chip-level parasitics.

Following chip-level parasitic extraction, the iso-voltage performance of both implementations is evaluated through sign-off analysis (Table~\ref{tab:results_MAC_table}). The shorter total wire length of the CFET implementation enables a higher frequency, resulting in a throughput of \qty{1.81}{TOPS}, compared with \qty{1.39}{TOPS} for the NSFET implementation. However, this performance improvement is accompanied by a \qty{75}{\percent} increase in power consumption. Combined with the smaller chip area, this leads to an increased power density from \qty{70}{\watt\per\centi\meter\squared} for the NSFET-based chip to \qty{162}{\watt\per\centi\meter\squared} for the CFET chip. Comparing these results with the pre-physical design results reveals that the increased chip-level performance mainly stems from improved routing, which outweighs the CFET cell delay penalty. This highlights the importance of system-technology co-evaluation.

Next, voltage-scaling sign-off analysis is performed. As shown in Fig.~\ref{fig:scaling}, the A7 CFET design operating at \qty{0.56}{\volt} achieves iso-frequency operation with the A10 NSFET design operating at \qty{1.0}{\volt}, while reducing the total power by \qty{68}{\percent}. Under iso-power operation, the CFET design provides a \qty{28.8}{\percent} frequency improvement. The NSFET design reaches a power density of \qty{148}{\watt\per\centi\meter\squared} at \qty{1.0}{\volt}, whereas the iso-frequency CFET design reaches only \qty{55}{\watt\per\centi\meter\squared} at \qty{0.56}{\volt}. Thus, the CFET power-density concern is substantially eliminated when the routed-chip timing advantage is traded-off with a lower operating voltage.

\subsection{BTI-Aging Results}

The PPA-derived iso-frequency power densities are subsequently converted into steady-state temperatures using multiphysics simulations, as shown in Fig.~\ref{fig:full_chip_thermal}. The resulting aging stress points are \qty{1.0}{\volt}/\qty{125}{\celsius} for A10 NSFET and \qty{0.56}{\volt}/\qty{62}{\celsius} for A7 CFET. The stress pairs are used in 10-year physics-based BTI simulations in Ginestra\texttrademark{}. As shown in Fig.~\ref{fig:deltavth}, $\Delta V_{TH}$ decreases from \qty{61}{\milli\volt} for NSFET to \qty{37}{\milli\volt} for CFET. The extracted $\Delta V_{TH}$ are then propagated to the circuit level through aging-aware standard-cell libraries, which we characterized. Using aging-aware standard-cell libraries and chip-level sign-off analysis, we demonstrate a \qty{53}{\percent} reduction in the accelerator aging timing guardband at the CFET stress point relative to the NSFET stress point (Fig.~\ref{fig:guardband}).


Overall, the improved routing efficiency of the CFET implementation provides a substantial speed advantage that can be traded for lower-voltage operation, thereby improving reliability while maintaining the baseline NSFET performance.

\section{Conclusion}
We presented parasitic-aware, thermal-aware, and aging-aware system--technology co-evaluation for A7 CFET and A10 NSFET designs. The
underlying technologies are coupled with automated cell layout generation, 3D parasitic extraction, library characterization, routed-chip implementation, multiphysics thermal simulation, and BTI-aware timing analysis. 
Our work reveals how parasitic RCs and design-rule scaling lead to three coupled effects that are difficult to distinguish using a single abstraction level. \textbf{\Circled{1}} CFET vertical stacking improves density but increases cell parasitics. \textbf{\Circled{2}} The smaller routed design reduces wire length, hence chip-level parasitics decrease - opposite to cell-level parasitics. \textbf{\Circled{3}} The gain in performance, due to lesser routing, enables lower-voltage iso-frequency operation, which mitigates thermal and aging effects. 
Results demonstrate that CFET should be assessed from cell parasitics to chip reliability: cell-level penalties can co-exist with system-level gains when routed-chip compactness enables lower-voltage iso-frequency operation.


\section*{Acknowledgment}

We thank S. Shahin, S. Deshwal, A. Mema, S. Niftaliyev, and M. Wei from TUM for their help in device-/system-level simulations. We also thank Applied Materials for providing access to SLiC™ as well as Gaurav Thareja from Applied Materials for his very valuable support and discussions.

\printbibliography

\end{document}